\documentclass[twocolumn,aps,prl,superscriptaddress,showpacs]{revtex4}
\usepackage{amsmath}
\usepackage{hyperref}
\usepackage{epsfig}
\usepackage{epsf}
\usepackage{array}
\usepackage{color}
\usepackage{graphicx}
\usepackage{epstopdf}
\usepackage{psfrag}
\usepackage{float}

\begin{document}
\bibliographystyle{apsrev}

\graphicspath{"./figures/"}

\title{Superconductivity and the Pseudogap in the Two-Dimensional Hubbard Model}

\author{Emanuel Gull}
\affiliation{Department of Physics, University of Michigan, Ann Arbor, MI 48109, USA}
\affiliation{Max Planck Institute for the Physics of Complex Systems, Dresden, Dresden 01187, Germany}
\author{Olivier Parcollet}
\affiliation{Institut de Physique Th{\'e}orique, CEA, IPhT, CNRS, URA 2306, F-91191 Gif-sur-Yvette, France}
\author{Andrew J. Millis}
\affiliation{Department of Physics, Columbia University, New York, New York 10027, USA}
\begin{abstract}
Recently developed numerical methods have enabled the explicit construction of the superconducting state of the Hubbard model of strongly correlated electrons in parameter regimes where the model also exhibits a pseudogap and a Mott insulating phase.   $d_{x^2-y^2}$ symmetry superconductivity is found to occur in proximity to the Mott insulator, but separated from it by a pseudogapped nonsuperconducting phase.  The superconducting transition temperature and order parameter amplitude are found to be maximal at the onset of the normal-state pseudogap. The emergence of superconductivity from the normal state pseudogap leads to a decrease in the excitation gap. All of these features are consistent with the observed behavior of the copper-oxide superconductors. \end{abstract}

\maketitle 

Lamellar perovskite-based copper oxide compounds display three remarkable properties: $d$-wave superconductivity with unprecedentedly high transition temperatures \cite{Bednorz86}, a nontrivial (``Mott'') insulating state \cite{Anderson87} and non-Fermi-liquid physics, most notably  a  ``pseudogap'' regime  in which the density of states is strongly suppressed in some parts of the Brillouin zone but not in others \cite{Hufner08}.  P. W. Anderson \cite{Anderson87} argued that these three classes of phenomena  have a common origin as strong-correlation effects understandable in terms of the two-dimensional repulsive Hubbard model, a minimal model of interacting electrons on a lattice with Hamiltonian
\begin{equation}
H = \sum_{p,\sigma} \left(\epsilon_{p} -\mu\right)
c^\dagger_{p,\sigma}c_{p,\sigma}+U\sum_i n_{i,\uparrow}n_{i,\downarrow},
\label{H}
\end{equation}  
where $\varepsilon_p=-2t(\cos p_x+\cos p_y)+4t'\cos p_x \cos p_y$  an electron dispersion and $U>0$  a local interaction which disfavors double occupancy of a site. 

In the years since Anderson's paper, the interplay of the pseudogap and superconductivity and the relation of both to the Hubbard model have been of central interest to condensed matter physicists. The existence of $d$-wave superconductivity in the Hubbard model has been demonstrated by perturbative analytic calculations \cite{Zanchi96} (later improved by renormalization group methods \cite{Halboth00,Raghu11})  and by numerics  \cite{Maier05_dwave,Scalapino06}.  The issue of the pseudogap has been more controversial.  It has been variously argued that  the pseudogap is a signature of unusual superconducting fluctuations \cite{Emery95,Engelbrecht98,Ong02}, of  a competing nonsuperconducting phase or regime  \cite{Hufner08,Taillefer09}, or of physics not contained in the Hubbard model \cite{Varma06}.  Theoretical determination of the interplay of the pseudogap and superconductivity in the Hubbard model is important in helping resolve this controversy, and will provide insight into the pseudogap phenomenon and into strongly correlated superconductivity more generally, but this requires access to intermediate or strong couplings for which perturbation theory is inadequate. %One reason the issue has remained unresolved is although many phenomenologically based theories have been studied, there is a lack of theoretical information concerning the superconducting properties of the Hubbard model itself. The existence of  $d_{x^2-y^2}$ superconductivity has been demonstrated by analytic calculations, originally perturbatively about suitably defined weak coupling limits \cite{Zanchi96} and later improved by renormalization group methods \cite{Halboth00,Raghu11}, but these calculations are valid in the weak coupling limit; numerics are apparently needed to obtain information about the strong coupling, low-doping regime where the pseudogap and superconductivity may compete. 

\begin{figure}[t]
\includegraphics[width=0.9\columnwidth]{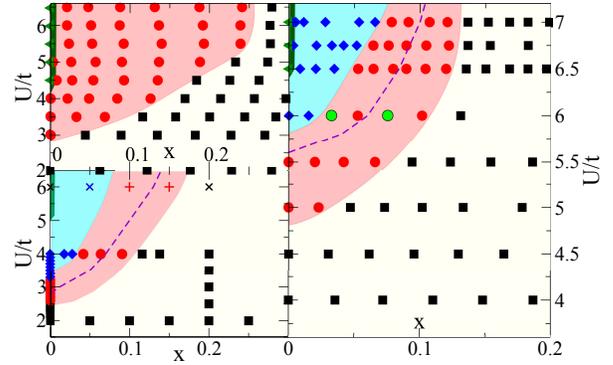}
\caption{Superconducting phase diagram of the two-dimensional Hubbard model in the plane of interaction strength $U$ and carrier concentration $x$ computed using the $8$-site (right panel), the  $4$-site  (left upper panel), and $16$-site (left lower panel) DCA dynamical mean field approximation at temperature $T=t/40$ with $t'/t=0$. Dashed line: location of the normal state pseudogap onset.  Circles  and shading (red online) indicate the superconducting region; squares (black online) and no shading the nonsuperconducting Fermi liquid; diamonds  and lighter shading (blue online) the nonsuperconducting pseudogap region; triangles and heavy solid line (dark green online) the Mott insulating region at $n=1$ and $U>U_c$. Open circles (light green online) denote the points analyzed in Fig.~\ref{FigSpec}. ``$\times$'' and ``$+$'' symbols denote 16-site data at $U/t=6$ extrapolated from a normal state solution, reproduced from \textcite{Yang11}.
}
\label{FigPD}
\end{figure}
The development of cluster dynamical mean field theory \cite{Maier05} has provided important nonperturbative information about the Hubbard model. Dynamical mean field theory approximates the electron self-energy in terms of a finite number of auxiliary functions determined from the solution of an $N$-site quantum impurity model and  becomes exact as  $N$ tends to infinity. In this Letter  we use dynamical mean field methods to   determine the interplay of superconductivity and the pseudogap in  the Hubbard model. This is challenging  because the theory of the superconducting  state involves both normal (N) and anomalous (A) components of the Green's function $G$ and self-energy $\Sigma$, leading to a doubling of the size of all matrices involved in the calculation, and hence to at least an eightfold increase in computational burden, which is further increased by the need to reach very low temperatures. 

We have constructed the superconducting state and studied its interplay with the pseudogap using clusters of   $N=4,8,16$  sites,  a size range found in previous work \cite{Gull10_clustercompare} to be large enough to distinguish generic $N\rightarrow\infty$  behavior from that specific to particular clusters. Specifics of our methods are given in the Supplemental Material; here we briefly note that a key aspect of our study is the use of  recently developed ``submatrix update'' numerical techniques \cite{Gull08_ctaux,Gull10_submatrix,Gull11_review} which enable access to couplings strong enough to produce a pseudogap at temperatures low enough to construct the superconducting state for cluster size $N$ large enough to reasonably represent the $N\rightarrow\infty$ limit.  Our key results are  that the pseudogap and superconductivity are competing phases and that, remarkably,  the onset of superconductivity within the pseudogap phase leads to a decrease in the excitation gap, in sharp contrast to conventional situations where the onset of superconductivity increases the gap.

Our analysis builds on previous dynamical mean field results. In pioneering papers Lichtenstein and Katsnelson \cite{Lichtenstein00} and Maier {\it et al.} \cite{Maier00} showed that  the $N=4$ cluster dynamical mean field approximation yielded $d_{x^2-y^2}$ superconductivity while subsequent studies of Maier and collaborators \cite{Maier05_dwave} on clusters with $N$ as large as  $26$  provided convincing evidence that the superconductivity found in the small cluster calculations is not an artifact, but rather is a property of the infinite cluster size limit, i.e. of the Hubbard model.
However, the studies of Ref.~\cite{Maier05_dwave} were restricted to to a modest interaction, $U = 4t$, too small to give a pseudogap, and to  relatively high temperatures, so that the the superconducting state was not constructed and  transition temperature was inferred from studies of the pair susceptibility.
Very recently Yang and collaborators \cite{Yang11} analyzed the pairing susceptibility for higher interaction strengths where a pseudogap occurred, but still did not construct the superconducting state. 

The pioneering work of Huscroft {\it et al.} \cite{Huscroft01} showed the existence of a normal-state pseudogap in the dynamical mean field approximation and many authors  (using mainly $N=4$ approximations) have studied its properties   \cite{Parcollet04,Civelli05,Kyung06,Macridin06,Maier06,Maier07,Maier07B,Chakraborty08,Kancharla08,Liebsch08,Koch08,Park08,Gull08_plaquette,Ferrero09,Stanescu06,Sakai09,Liebsch09,Vidhyadhiraja09,Sordi10} and several groups (still within  the 4-site approximation) have studied the interplay of superconductivity and the pseudogap \cite{Haule07,Maier08,Kancharla08,Civelli08,Civelli09B,Sordi12}. A key finding of the 4-site work, in contrast to the larger-cluster studies of Ref.~\cite{Yang11} is that superconductivity persists all the way to the Mott insulating boundary, leaving open the question whether it is the pseudogap per se, or simply Mott physics, which suppresses the superconductivity.

More recent developments \cite{Gull10_submatrix} have enabled researchers to access clusters large enough to obtain a reasonable picture of the $N\rightarrow\infty$ limit  \cite{Werner098site,Gull09_8site,Gull10_clustercompare,Khatami10,Yang11,Chen11,Sakai12}. It has been found \cite{Gull10_clustercompare} that in DCA clusters of size $N>4$ the Mott transition  is multistaged, with the  fully gapped Mott insulating state being separated from the Fermi liquid state by an intermediate phase, in which regions of momentum space near the $(0,\pi)/(\pi,0)$ point are gapped and regions of momentum space near $(\pm \pi/2,\pm \pi/2)$ are not.  By contrast, in most of the  $N=2,4$ calculations reported to date there is at half filling no intermediate phase separating the insulator and the Fermi liquid  \cite{Gull08_plaquette,Park08}, while if the insulator is destroyed  by doping an intermediate phase  with a suppressed, but nonzero, density of states is found  \cite{Gull08_plaquette,Park08,Sordi10}.  In this Letter we extend the new methodology to examine the properties of the superconducting state at $N$ large enough to properly represent the pseudogap.

%The right panel of Fig.~\ref{FigPD} presents the phase diagram Consistently across all clusters and in agreement with previous results \cite{Zanchi96,Halboth00,Maier05_dwave} we found that the Hubbard model exhibits  $d_{x^2-y^2}$-symmetry superconductivity, with a typical transition temperature $\sim t/40 \approx 100K$ (using a $t \approx 0.3eV$ representative of the CuO$_2$ superconductors). At the temperatures accessible to us, superconductivity occurs only for interactions larger than a critical value and dopings sufficiently close ($\lesssim 20\%$) to half filling, but persists up to the largest interactions ($U/t=8$) we have investigated.  

The right-hand panel of Fig.~\ref{FigPD} shows the phase diagram determined from a comprehensive survey of parameter space for the $N=8$ dynamical cluster approximation, which previous work \cite{Gull10_clustercompare} shows adequately represents the $N\rightarrow\infty$ normal state physics of the model. Studies of selected $U$ and doping values in the computationally much more expensive $N=16$ site cluster confirm (lower left panel) that the physics found for $N=8$ is generic. The scan of the phase diagram is conducted at temperature $T=t/40$ but checks of selected interaction and doping values at our lowest accessible temperature $T=t/60$ (see also Ref.~\cite{Gull12_energy}) indicate that lower temperatures do not bring significant changes (see Supplemental Material).   

$d_{x^2-y^2}$-symmetry superconductivity, with a typical transition temperature $\sim t/40 \approx 100K$ (using a $t \approx 0.3eV$ representative of the CuO$_2$ superconductors) occurs in a band of interaction strength and density, vanishing if interaction or doping are tuned too far away from the insulating state but separated from the Mott insulator by a region of pseudogapped but nonsuperconducting states. This result, previously inferred from extrapolation of the pairing susceptibility \cite{Yang11} at high temperature, is here confirmed.  The onset of the normal state pseudogap (dashed line) corresponds to the maximum in the superconducting order parameter (see Supplemental Material) and to the maximum in transition temperature (see below). The inset of Fig. 2, Supplemental Material, shows that the superconducting region remains separated from the pseudogap even as $T\rightarrow 0$.

 %We shall show below that in the superconducting/pseudogap regime (above and to the left of the dashed line) the superconductivity is highly anomalous: in contrast to the usual situation, the transition to superconductivity from the pseudogap state results in a {\em decrease} of the excitation gap.

The upper left panel shows that the situation is  different in the $N=4$ approximation. In this case, superconductivity extends all the way to the boundary of the Mott phase, as has previously been found  \cite{Jarrell01,Civelli08,Civelli09B,Sordi12}. We believe that the difference arises because in the $8-$ and $16-$ site cluster approximations the pseudogap leads at $T=0$ to a complete suppression of the density of states in the momentum region ($0,\pi$) important for superconductivity; in the  4-site approximations the pseudogap produces a density of states which is suppressed relative to the Fermi liquid, but is still nonvanishing in the regions important for superconductivity (see, e.g. Fig. ~[3] of Ref.~\cite{Gull08_plaquette} or Fig.~[2] of Ref.~\cite{Sordi12}).  Variational Monte Carlo studies \cite{Gros88,Yokoyama88,Becca00,Paramekanti01,Paramekanti04,Yokoyama04,Ogata12a,Ogata12b} also do not find an intermediate nonsuperconducting phase; the difference may have to do with the ability of the variational wave functions to represent the physics of the pseudogap but this issue demands further research.

\begin{figure}[Tb]
\includegraphics[width=0.9\columnwidth]{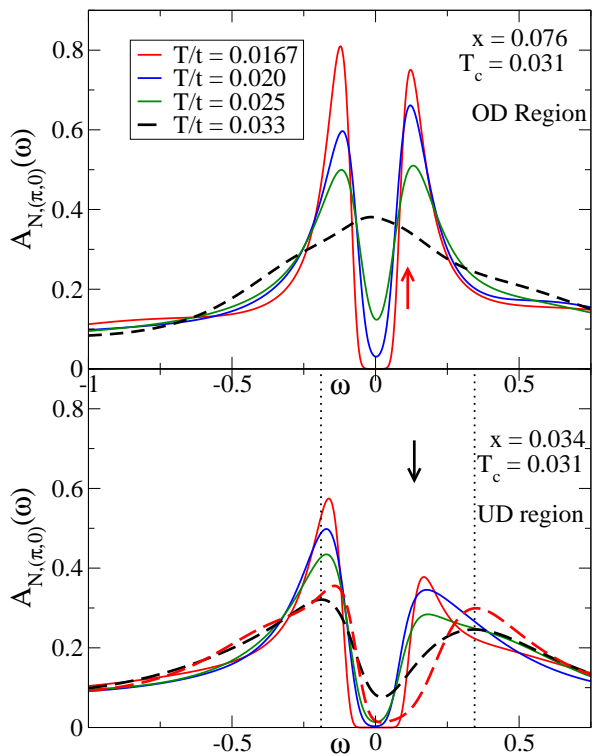}
\caption{Analytically continued  spectral function computed at $U=6t$ for the antinodal sector showing temperature evolution of gap structure for a typical optimally doped or overdoped  state ($x=0.076$, upper panel) and underdoped  pseudogap state ($x=0.034$, lower panel). Solid lines: superconducting spectral function.  
Heavy dashed lines: normal state spectral function, obtained for $T=t/30$. Light dashed line (lower panel): normal state density of states at $T=t/60$ obtained by  suppressing superconductivity. Arrows mark spectral function maxima used to determine superconducting gap size $\Delta$. Dotted lines:  pseudogap energy  at $T=t/30$ obtained from the maximum in the spectral function.
%Dashed red line: normal state spectral function at $T=t/60$ computed by suppressing superconducting order.
% Inset: Dashed lines: Difference of running integral $I=\int_{-\omega}^\omega A(x) dx$ between superconducting $(T=t/60)$ and normal $(T=t/30)$ state in the pseudogap (UD) and overdoped (OD) regime showing that spectral weight in superconducting coherence peaks is drawn from a very wide range of frequencies in the UD case, $|\omega| \gg 4\Delta$, and from a  region $|\omega| \lesssim 4\Delta$ in the overdoped case. Dash-dotted lines: $4\Delta$.
}
\label{FigSpec}
\end{figure}

Figure~\ref{FigSpec} presents the frequency and temperature dependence of the density of states. The upper  panel  shows spectra representative of dopings higher than, or interactions weaker than, the values which maximize $T_c$, so that superconductivity emerges from a relatively conventional normal state. The spectra are consistent with expectations from standard theory \cite{Bardeen57}: the onset of superconductivity is associated with a suppression of density of states at low frequency and with the formation of density of states (``coherence'') peaks. We define the superconducting gap $\Delta$ as half of the peak to peak distance. The  area in the coherence peaks comes mainly from the states removed at $|\omega|<\Delta$. The gap amplitude develops very rapidly with temperature: only at the temperature closest to $T_c$ is the peak to peak splitting appreciably different from its value at the lowest $T$. 

The situation is quite different when superconductivity emerges from the pseudogap regime. Representative spectra are shown in the lower panel of Fig.~\ref{FigSpec}. The  normal state pseudogap is visible at $T>T_c$ as a suppression of the density of states at low frequencies with a broad gap structure at higher frequencies. The $T<T_c$ normal state density of states (obtained by suppressing superconductivity) displays essentially the same behavior.  The development of superconductivity is characterized by the formation of coherence peaks at energies {\em below} the pseudogap, i.e. by a {\em decrease} in gap magnitude as the superconducting state is entered.    This behavior is consistent with recent experimental reports ~\cite{He11} that in underdoped cuprates the emergence of  superconductivity  out of the pseudogap regime is associated with the formation of new states at energies lower than  the pseudogap energy and that the superconducting gap is tied to the pseudogap. Furthermore, most (typically more than $50\%$) of the spectral weight in the coherence peak is drawn from frequencies greater than $\Delta$.

\begin{figure}[Tb]
\includegraphics[width=0.9\columnwidth]{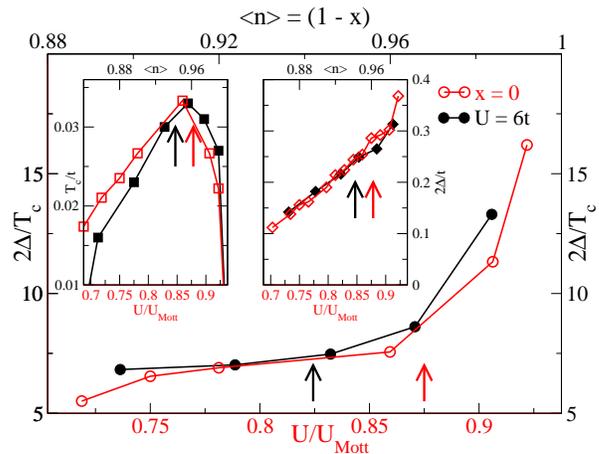}
\caption{Gap to transition temperature ratio $2\Delta/T_c$ (main panel) computed using the $8$-site DCA approximation both by varying $U$ at  $x = 0$ (open symbols, lower axis, red color, $U_\text{Mott}=6.4t$) and by varying $x$ for $U=6t$  (filled symbols, upper axis, black). Gap defined as peak to peak distance in analytically continued spectral function. Left inset, squares: doping and interaction dependence of transition temperature for same parameters, showing superconducting dome. Right inset, diamonds: doping and interaction dependence of gap $2\Delta/t$. Arrows: onset of normal state pseudogap.}
\label{FigGapVsTc}
\end{figure}

Figure~\ref{FigGapVsTc} presents the superconducting transition temperature  determined as described in the Supplemental Material, as well as the gap values obtained as described above. Similar to the anomalous expectation value (inset, Fig. ~\ref{FigPD}), the transition temperature has a domelike behavior, with the highest transition temperature occurring near the onset of the normal state pseudogap (insets of Fig.~\ref{FigGapVsTc}), whereas the gap monotonically increases from high to low doping or low to high interaction. 
%
%by performing a linear extrapolation of the square of the equal-time anomalous Green's function $G_{A,(\pi,0)}$, using the three highest temperatures at which %$G_A>0.1$.  We find . We cross-checked the transition temperature estimates by computing the temperature at which the inverse of the normal state pairing %susceptibility vanishes.
%
%We obtained the superconducting gap by analytically continuing the $G_N(\tau)$ and reading off the distance between the quasiparticle peaks. We cross-checked the gap values by analytically continuing the self-energies  (see Supplemental Material). 
%In contrast to the nonmonotonic behavior of the transition temperature vs doping or interaction curves, the  superconducting gap  increases monotonically 
%from high to low doping or low to high interaction.%, and indeed varies by a factor of $\sim 2$ over the superconducting range. 
We find $2\Delta/T_c \sim 7.5-8$ in the region outside the pseudogap   and becoming rapidly larger within the pseudogap regime as the endpoint of the superconducting regime is approached, consistent with dynamical mean field calculations based  on 4-site clusters  \cite{Haule07,Maier08,Kancharla08,Civelli08,Civelli09B,Sordi12}.    In interpreting the numerical value of the gap it is important to note that the DCA procedure, which averages over an entire momentum sector, places the peak at a somewhat higher energy than the true minimum excitation energy. This difference does not affect the trends of primary interest here.  

\begin{figure}[htbp]
\includegraphics[width=0.9\columnwidth]{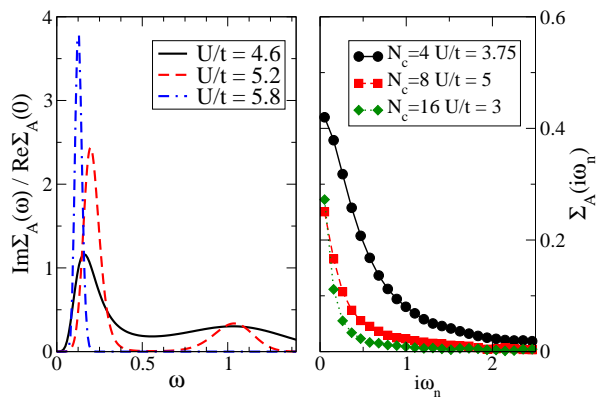}
\caption{Left panel: Imaginary part of self-energy calculated for particle-hole symmetric Hubbard model at $n=1,$  $T=t/60$ and indicated interaction strengths by directly continuing the Matsubara axis self-energy.  Right panel: comparison of frequency dependence of anomalous self-energy for $4-,$ $8-,$ and $16-$site cluster approximations computed at $n=1$, $T=t/60,$ and $U$ values indicated.}
\label{sigs}
\end{figure}

Further insight into the superconductivity may be obtained from the imaginary part of the real-axis anomalous self-energy obtained by maximum entropy analytical continuation as described in the Supplemental Material and shown in the left panel of  Fig.~\ref{sigs}. In standard phonon-mediated superconductivity Im$\Sigma_A$ is peaked at frequencies associated with the phonons \cite{Scalapino66}.  At the weaker coupling $U=4.6$ Im$\Sigma_A$ is spread over a range of frequencies  up to somewhat larger than $\omega=t$, possibly consistent with a spin fluctuation origin of superconductivity but  as the coupling is increased the weight shifts dramatically to lower frequencies, and for the strongest couplings essentially all of the weight is concentrated in a very low frequency peak. This  strong coupling behavior is highly unusual, and requires further analysis. We also remark that  our $8-$ and $16-$ site cluster calculations do not show  evidence for the contribution from higher frequency ($\omega \sim U$) scales reported by Ref.~\cite{Maier08} (see also ~\cite{Kyung09}). This conclusion is not dependent on analytical continuation: a contribution along the lines of that reported in Ref.~\cite{Maier08} would lead to a Matsubara-axis anomalous self-energy which at $\omega\sim 2t$  would be  $\sim 20\%$ of its zero frequency value. 
As can be seen from the right-hand panel of Fig.~\ref{sigs} while in the 4-site cluster the Matsubara axis anomalous $\Sigma$ function may be different from zero for $\omega \sim 2t$, for the larger clusters it clearly has decayed to zero for $\omega \gtrsim 2t$.
%As can be seen from the right-hand panel of Fig.~\ref{sigs} while in the 4-site cluster the Matsubara axis anomalous Green function may tend to a nonzero limit for $\omega\sim 2t$, for the larger clusters $\Sigma_A(i\omega_n)$ falls rapidly to zero as the frequency is increased beyond $\omega\sim t$.  

%Summary
%\paragraph*{Conclusion}
%In summary, previous work using a variety of   approaches \cite{Zanchi96, Maier05_dwave} has shown that the Hubbard model is superconducting, while previous 4-site cluster dynamical mean field studies   \cite{Lichtenstein00,Maier00,Haule07,Maier08,Kancharla08,Civelli08,Civelli09B,Sordi12} have indicated that superconductivity is interesting, being characterized in particular by an anomalous self-energy which increases to very large values as the insulating phase is approached.   In this paper we have shown, robustly over a range of cluster sizes, interaction strengths, and carrier concentrations, that the superconducting phase occurs within a superconducting dome bounded on one side by the weak coupling/large doping regime and on the other by proximity to the Mott insulating phase. In all clusters except the 4-site ones,  superconductivity is however  separated from the Mott insulator by a pseudogapped but nonsuperconducting phase. When superconductivity emerges from the pseudogap regime it fundamentally reconstructs the  density of states, leading in particular to new states within the pseudogap.

In summary, we have constructed the superconducting phase and analyzed its competition with the pseudogap. We find, robustly over a range of cluster sizes, interaction strengths and carrier concentrations, that in the Hubbard model the superconducting and pseudogap phases compete. The competition is manifested by the presence of a pseudogapped but nonsuperconducting phase close to the Mott insulator and by a dramatic change in the density of states, in particular a decrease of the gap size when  superconductivity emerges from the pseudogap state.  In addition, we find that when superconductivity and the pseudogap coexist, the superconductivity is anomalous, with the imaginary part of the  self energies characterized by a sharp large amplitude pole at an energy near zero.

Our  results open up  important new directions for research.  
%Our  methods are restricted (at least in the present state of development) to models with local density-density interactions and to situations where the fermionic sign problem, which gets exponentially worse with increasing cluster size and decreasing $T$, is not too severe. In the context of the two-dimensional Hubbard model the practical restriction is 
For the two-dimensional Hubbard model fermion sign and matrix size issues restrict us in practice to $N_c \lesssim 16$ and interaction $U \lesssim 7$.   These values are large enough to enable access to the doped Mott phase while accessing large enough cluster sizes to obtain reasonable insight into the infinite cluster size limit.  Even given these constraints,  understanding the anomalous frequency dependence of the anomalous self-energy at strong coupling  and further investigation  of the interplay between the pseudogap and the superconducting gap, and investigation of two-particle (e.g., Raman) spectra are feasible. In particular the striking similarity between the physical behaviors of the doping-driven and interaction-driven transitions shown in Fig.~\ref{FigGapVsTc} suggests that the computationally simpler particle-hole symmetric case will provide valuable generally valid information.  Going beyond the particle-hole symmetric case, investigations of the effect of second neighbor coupling are important to determine the factors optimizing $T_c$.  Also, a significant difference between our calculations and experiment is that we find a larger anomalous Green function on the electron doped side. Inclusion of long-ranged antiferromagnetism and also extension of our results to the `three-band' copper oxide models is needed to understand these issues further.

We thank Thomas Maier and Michel Ferrero for helpful discussions. A.~J.~M. and E.~G. were supported by NSF-DMR-1006282, O.~P. by the ERC grant ``MottMetals''. This research used resources of GENCI-CCRT (Grant No. 2011- t2011056112) and of the National Energy Research Scientific Computing Center, which is supported by the Office of Science of the U.S. Department of Energy under Contract No. DE-AC02-05CH11231. A portion of this research was conducted at the Center for Nanophase Materials Sciences, which is sponsored at Oak Ridge National Laboratory by the Office of Basic Energy Sciences, U.S. Department of Energy. 
All authors contributed to the design of the study, creation of  the algorithm, data analysis, and manuscript preparation. The computer code was written and simulation data were produced by E.G.
\bibliography{refs_shortened}

\end{document}

\author{Emanuel Gull}
\affiliation{Department of Physics, University of Michigan, Ann Arbor, MI 48109, USA}
\affiliation{Max Planck Institute for the Physics of Complex Systems, Dresden, Germany}
\author{Olivier Parcollet}
\affiliation{Institut de Physique Th{\'e}orique, CEA, IPhT, CNRS, URA 2306, F-91191 Gif-sur-Yvette, France}
\author{Andrew J. Millis}
\affiliation{Department of Physics, Columbia University, New York, New York 10027, USA}
\title{Supplementary Material: Superconductivity and the Pseudogap in the two-dimensional Hubbard model}

\pacs{
71.27.+a,%Strongly correlated electron systems; heavy fermions 
71.28.+d,%Narrow-band systems; intermediate-valence solids
78.30.-j,%Raman spectroscopy
74.72.Kf,%Cuprates/Pseudogap regime 
}

\maketitle

%\appendix
%%%%\section{Sign Problem}
%%%%While critical slowing down and the $O((N\beta U)^3)$ scaling of the matrix updates make results at low temperatures, large interaction strengths, large cluster sizes, and near a phase boundary difficult to obtain, the ultimate limitation is given by the fermionic sign problem. The sign problem is strongly dependent on the system studied, but in general is much less severe for impurity problems than for lattice problems of comparable size and with comparable interaction parameters.
%%%%
%%%%Outside of the superconducting phase our algorithm simplifies to the normal state CT-AUX algorithm and has the same sign problem. We find that the superconducting solution, where it exists, always has a less severe sign problem than the normal state solution at the same temperature. This increase in sign is not big enough that significantly larger interaction strengths, cluster sizes, or lower temperatures become accessible.
%%%%The increase of the average sign is consistent with our experience in other systems where often the average sign raises as soon as a gap is developed. 
%%%%
%%%%At particle-hole symmetry ($t'=0, n=0.5$), there is no sign problem. As in the normal state, systems with a next-nearest neighbor hopping have a smaller average sign.

%%%Fig.~\ref{FigSign} shows representative data for the superconducting and pseudogap regions on the eight-site cluster.
\section{Numerical procedure}
%\subsection{Method}
We study the two dimensional Hubbard model (Eq.~1 of main text) using the dynamical cluster approximation (DCA) version of cluster dynamical mean field theory  \cite{Maier05}, with our recent implementation of a Nambu (superconducting-state) version of the numerically exact continuous-time auxiliary field \cite{Gull08_ctaux,Gull11_review} quantum impurity solver with submatrix updates \cite{Gull10_submatrix} based on the  open source ALPS \cite{ALPS20,ALPS,ALPS_DMFT} libraries. We study cluster sizes $N=4,8,16$ using the standard cluster tilings.

To describe the DMFT self-consistency process it is convenient to use the Nambu matrix notation introducing Pauli matrices $\tau_{0,1,2,3}$ in particle-hole space so that the Green function $\bf G$, self energy $\bf \Sigma$ and mean field function ${\bf \mathcal{G}}_0^{-1}$ have %Eq.~\ref{ksum} is inverted in Nambu space and 
%A mean field function ${\bf \mathcal{G}}_0^{-1}(K,i\omega_n)$ with  
normal (N) and anomalous (A) components (e.g. $\mathbf{ \Sigma}=\mathbf{\tau}_0\Sigma_N+\mathbf{\tau}_1\Sigma_A$) \cite{Lichtenstein00}.  The mean field function is determined from the Nambu matrix equation
\begin{equation}
{\bf \mathcal{G}}_0^{-1}(K,i\omega_n) = {\bf \Sigma}(K,i\omega_n) + \frac{N}{N_{latt}} \left[\sum_{k\in K} {\bf G}_{latt}(k,i\omega_n)\right]^{-1}
\end{equation}
Here $K$ labels momentum sectors,   $N_{latt}\rightarrow\infty$ the number of sites in the lattice and the $\sum_{k\in K}$ is over the $N_{lattice}/N$ points in sector $K$ of the Brillouin zone of a square lattice of $N_{latt}$ sites and periodic boundary conditions.
\begin{equation}
{\bf G}_{latt}(k,i\omega_n) = \left[i\omega_n {\tau}_0 + (\mu - \epsilon_k){\bf \tau}_3 -{\bf \Sigma}(K,i\omega_n)\right]^{-1}
\end{equation}
 and  $\bf \Sigma$ is obtained from a solution of the corresponding impurity model. 
%Here $K$ labels momentum sectors, ${\bf \tau}_{0,1,3}$ are the usual Pauli matrices in Nambu (particle-hole) space, and $\bf G$ as well as $\bf \Sigma$ and ${\bf \mathcal{G}}_0^{-1}$ have  normal and anomalous components.

The extension to superconductivity means that all matrices are twice as large as in a normal state computation at the same temperature. The doubling of matrix size means that submatrix update techniques \cite{Gull10_submatrix} are crucial for accessing the  large interaction strengths, low temperatures and large clusters needed in this study. Critical slowing down means that typically more than $50$ iterations are needed to converge near the superconducting phase transition.

%Critical slowing down means that typically more than $50$ iterations are needed to converge near the superconducting phase transition. %Fast update techniques \cite{Gull10_submatrix} are crucial for accessing large interaction strengths and low temperatures. The onset of superconductivity drastically changes the low energy electronic structure. One consequence is that in some parameter regimes, care is required in converging to the superconducting state.

%\subsection{Convergence procedure} 
Particularly in the pseudogap regime, the onset of superconductivity drastically changes the low energy electronic structure. A consequence is that care is required in converging to the superconducting state. Beginning a self-consistency loop by  adding a superconducting component to a converged non-superconducting solution leads to very slow convergence. %The choice of initial seed for the calculation is particularly important in the pseudogap regime, where the superconducting and normal solutions are very different, so that  adding a superconducting component to a converged non-superconducting solutions leads to very slow convergence. 
We have found that the most stable procedure is to begin at a relatively high temperature (e.g. $\beta = 10/t$) and introduce a pairing field $\eta_1(k) = \eta_1 \phi_k$ via the replacement ${\bf G}(k,i\omega_n;\eta_1) = \left[ i\omega_n {\bf \tau}_0 + (\mu - \epsilon_k){\bf \tau}_3 + \eta_1(k){\bf \tau}_1 -{\bf \Sigma}\right]^{-1}$, with e.g. $\phi_k=\cos k_x - \cos k_y$ for $d$-wave superconductivity, and $\eta_1$ typically 0.1t. Retaining the pairing field we obtain converged solutions ${\bf G}(K,i\omega_n;\eta_1)$ first at the initial temperature, then, using the solution at the initial temperature as a seed, at the desired range of lower temperatures. We remark that the sign problem for large $\eta_1$ is much less severe than at $\eta=0$, so these computations are not inordinately expensive. 

Then, at each temperature, using the converged ${\bf G}(K,i\omega_n;\eta_1)$ as a seed, we set $\eta_1=0$ in the self-consistency condition and continue iterating until convergence is reached. At selected points we check  the solution by taking the putatively converged self energy, dividing  the anomalous part by a large number (typically $20$), and verifying that under further iterations the solution converges back to the one previously found.

%Our codes are based on the open source ALPS \cite{ALPS20,ALPS,ALPS_DMFT} libraries.
\section{Phase Diagram}

\begin{figure}[htbp]
\includegraphics[width=0.8\columnwidth]{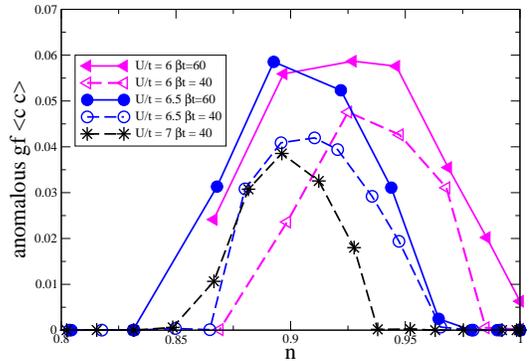}
\caption{Order parameter $\langle cc\rangle$ as a function of filling $n$ at $\beta t=60$ (solid lines and filled symbols) and $\beta t=40$ (dashed lines and open symbols), on an eight-site cluster, for interaction strengths indicated.}
\label{newcc}
\end{figure}
\begin{figure}[htbp]
\includegraphics[width=0.8\columnwidth]{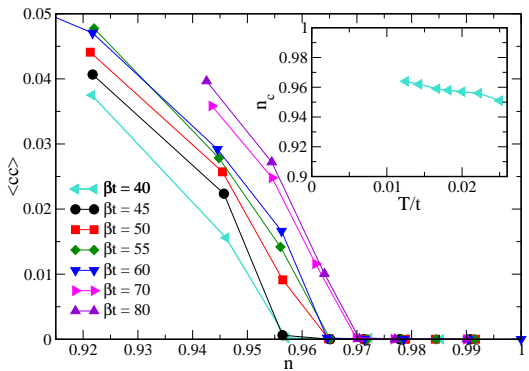}
\caption{Order parameter $\langle cc\rangle$ as a function of filling $n$ and $U/t = 6.5$ for various temperatures. Inset: scaling of the critical density as a function of temperature. Error bars are much smaller than symbol size.}
\label{cc_vs_T}
\end{figure}
\begin{figure}[htbp]
\includegraphics[width=0.8\columnwidth]{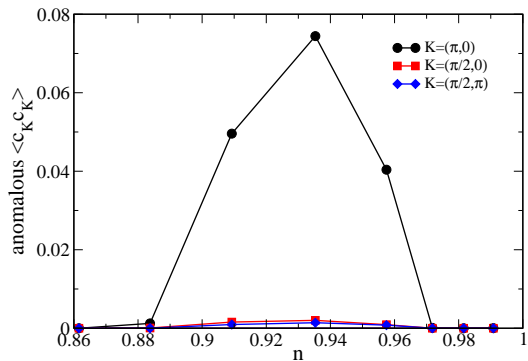}
\caption{Order parameter $\langle cc\rangle$ as a function of filling $n$ at $U/t = 4$ and $\beta t=50$ on a sixteen-site cluster, as a function of doping. Shown are three nonzero momentum sectors, with the main contribution coming from the $K=(\pi,0)$ sector.}
\label{sixteendoping}
\end{figure}

The location of the normal state pseudogap was determined  from the magnitude and temperature dependence of $G(\tau=\beta/2)$ as described in Ref.~\cite{Gull10_clustercompare}. The boundaries of the superconducting region were determined from computations of $G_A(0)=\langle c_{K\uparrow}(\tau)c_{K\downarrow}(\tau)\rangle$ at  temperature $T=t/40$ and the results were spot-checked against computations at $T=\beta/60$.  An example is presented here in Figures~\ref{newcc}, \ref{cc_vs_T} (8-site) and \ref{sixteendoping} (16-site).  At $U=6t$ (less than the $U_c\approx 6.4$ needed to produce a Mott insulator at $n=1$) the superconductivity extends all the way to $x=1$. As  $U$ is increased to $U=6.5 > U_c$ the superconducting phase clearly pulls away from $n=1$, with no significant difference in low-x endpoint between $\beta=40$ and $\beta=60$.  

This is further illustrated in Fig.~\ref{cc_vs_T}, which shows a high-resolution calculation of the low-doping end of the dome as a function of temperature. The inset shows the scaling of the critical filling as a function of temperature, for temperatures down to $\beta t=80$. The data suggests that the zero-temperature superconducting dome does not extend to half filling, in contradiction to calculations on smaller clusters and variational Monte Carlo calculations.

Figure~\ref{sixteendoping} shows the doping transition on a sixteen-site cluster at $\beta t=50$, exhibiting the superconducting dome separated from half-filling by a pseudogap state. Shown are also two additional momentum sectors $K=(\pi/2,\pi)$ and $K=(\pi/2,0)$, which show some, albeit relatively small, anomalous component of the self-energy.

\section{Transition temperatures and gap magnitudes}
\begin{figure}[htbp]
\includegraphics[width=0.9\columnwidth]{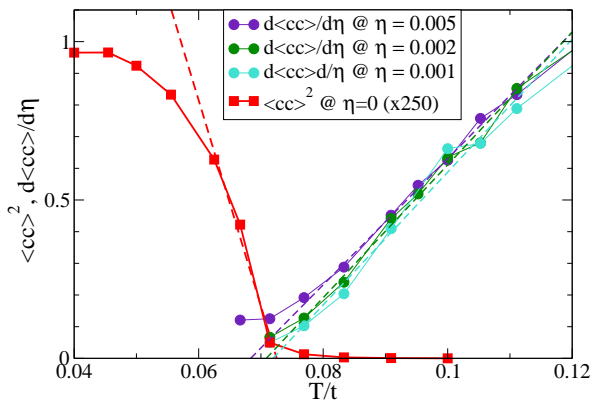}
\caption{$T_c$ obtained by extrapolation from $T>T_c$ and $T<T_c$, for a four-site cluster at $U/t=8$ and $\mu/t = -2$.  Filled squares (red online): $\langle cc \rangle^2$. Heavy dashed line (red online): Extrapolation using three  highest temperatures at which $<cc>^2>0.01$.  Circles: $d\langle cc\rangle/d\eta$ obtained from finite differences for $\eta=0.005$ to $\eta = 0.001$. Dashed light lines: Linear fit through $d\langle cc\rangle/d\eta$.}
\label{Tc_above_below}
\end{figure}

We estimated the transition temperature by performing a linear extrapolation of the square of the equal-time anomalous Green's function $G_{A,(\pi,0)}$, using the three highest temperatures at which $G_A>0.1$.  We cross-checked the transition temperature estimates by computing the temperature at which the inverse of the normal state pairing susceptibility (determined from the derivative of the anomalous expectation value with respect to an externally imposed pairing amplitude $\eta$) vanishes. An example is shown in Fig.~\ref{Tc_above_below}.

We obtained the superconducting gap by analytically continuing the $G_N(\tau)$ and reading off the distance between the quasiparticle peaks. We adopt this criterion because it is well defined and easy to verify and reproduce but we note that because the DCA procedure averages over a range of energies the resulting energy is larger than the minimum excitation energy in the system. 

\begin{figure}[htbp]
\includegraphics[width=0.9\columnwidth]{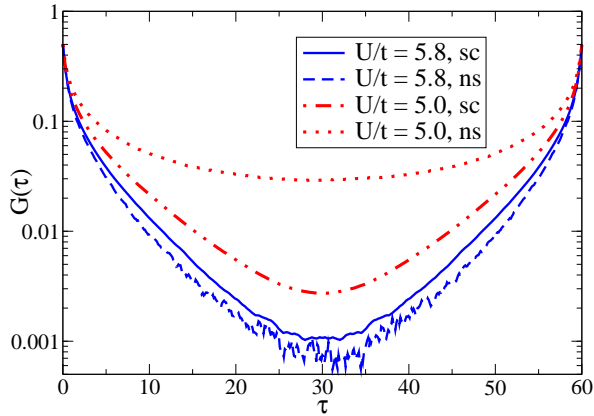}
\caption{Imaginary time Green function (normal, local component) directly measured in our quantum Monte Carlo calculation for normal states (dashed lines) and superconducting states (solid lines) for $n=1$ and   $U=5.0t$ (upper traces, red online) and $U=5.8t$ (lower traces, blue online). }
\label{Gtaufig}
\end{figure}

The crucial finding, namely that the gap is smaller in the superconducting than in the pseudogap state may also  be seen directly from our imaginary time data. Eq. ~\ref{spectra} implies that
\begin{equation}
\beta G\left(\tau=\frac{\beta}{2}\right)=\int \frac{d\omega}{\pi} \frac{A(\omega)}{4 \cosh\frac{\beta \omega}{2}}
\label{betaGbetaover2}
\end{equation}
In a system with a gap $\Delta$ Eq.~\ref{betaGbetaover2} implies $G(\tau=\beta/2)\sim e^{-\Delta/2T}$. Fig.~\ref{Gtaufig} shows that at $\beta=60$ and $U=5.8t$ the $G(\tau)$ computed with superconductivity suppressed indeed lies lower than the $G(\tau)$ computed in the superconducting state, indicating that the normal state has a larger gap than  superconducting state.

\section{Analytic Continuation}

\subsection{General observations}

We perform quantum Monte Carlo calculations, obtaining Green functions and self energies as functions of imaginary time. Real frequency information such as densities of states is obtained by inverting the relation between the imaginary-time Green's function ${\bf G}$ or self energy ${\mathbf \Sigma}$  and the associated spectral function $A$ (for ${\mathbf G}$) or  ${\mathbf \Sigma}^{''}$ (for ${\mathbf \Sigma}$). The relation is
\begin{equation}
{\bf G}(\tau) = \int_{-\infty}^\infty d\omega \frac{e^{-\tau\omega}}{1+e^{-\beta \omega}} {\bf A}(\omega).
\label{spectra}
\end{equation}
The inversion of  Eq.~\ref{spectra} is an ill-posed problem, because the kernel in Eq.~\ref{spectra} has many very small eigenvalues, so its inverse has many large ones, implying that small (statistical) fluctuations in the input data $G$ cause large fluctuations in $A$. To find a solution we  employ the maximum entropy continuation method \cite{Jarrell96}. 

Accurate knowledge of errors in the input data is crucial for a reliable continuation. We estimated these errors from a jackknife procedure  applied to  eight consecutive iterations of the converged solution. We assumed the covariance matrix to be diagonal in frequency, based on earlier work in the normal state \cite{Lin10}. We assessed the quality of our continuations by verifying that the back-continuation Eq.~\ref{spectra} did not contain systematic deviations from $G$ within the errors bars of the data.  Further confidence comes from the fact that the results do not change significantly as the precision is increased and that the change in the spectral functions is gradual and systematic as parameters ($U, T, \langle n \rangle$) are varied.  In particular our estimates of energy gaps and of densities of states at frequencies up to $\omega \sim 0.5t$ are robust to choices of model function and input data, and show reasonable trends between calculations. However, uncertainties remain in the continued quantities. Data near $\omega=0$ are in general more reliable and reproducible than data at high frequencies, and in particular  the continuations  contain small amplitude long-period oscillations which account for part of the differences seen between curves in Fig. 3 of the main text at $\omega\gtrsim 0.5t$.

\subsection{Continuation of self energies, superconducting state}

%Our previous experience \cite{Wang09gap} is that in situations where a gap is present,  the gap may be more accurately estimated by continuing the self energy and then reconstructing the Green function from the Dyson equation.  
In principle continuation of the anomalous self energy proceeds from Eq.~\ref{spectra} with anomalous spectral function $\Sigma_A^{''}$  used in place of $A$ and the Matsubara $\Sigma_A$ instead of $G$, but complications arise.  

First, $\Sigma_A^{''}$ is an odd function of frequency, so the issue of normalization must be handled differently. We rewrite the Kramers-Kronig relation in terms of $S=\Sigma_A^{''}(\omega)/\omega$ as
\begin{equation}
\Sigma_A(i\omega_n)=\int \frac{dx}{\pi}\frac{xS(x)}{i\omega_n-x}.
\label{continueanom}
\end{equation}
As Eq.~\ref{continueanom} indicates, $S$ is normalized to the zero frequency value of the anomalous self energy, which we obtain by a quadratic extrapolation of $1/\Sigma_A(i\omega_n)$ to $\omega_n=0$. A very accurate extrapolation is needed because an incorrect normalization will lead to spurious features in the continued function, either very near zero or at high frequency. 

Second, $\Sigma_A^{"}(\omega>0)$ need not be positive definite. It arises from an off-diagonal term in a Green function so the mathematical expression does not have the form of an amplitude squared. A well-known physical example is the conventional phonon-mediated superconductor in the presence of a Coulomb repulsion \cite{Scalapino66}. The familiar change in sign of the real part of the gap function at frequencies above the phonon energies implies mathematically a negative contribution to $\Sigma_A^{"}$ at high frequencies (typically of the order of the Coulomb scale). In the d-wave case of interest here the spatial symmetry of the cooper-pair wave function means that the two members of a Cooper pair have no probability to be on the same site, so the standard repulsive contributions do not operate and a frequency dependent sign change is not expected, but we know of no general argument which would rule out a sign change.

%Third, and of considerable practical importance, as can be seen from Fig. 4 of the main text or Fig.~\ref{sigs}, in the strong coupling limit the normal and anomalous terms in the self energy of the superconducting states are dominated by low frequency weakly damped poles, of comparable weight and position. These poles combine nontrivially in the reconstruction of the Green function and if $\Sigma_{N,A}$ are continued independently,  small differences in the position and strengths of these poles, arising from continuation errors, can lead to unphysical behavior when $G$ is reconstructed from $\Sigma$

\begin{figure}[t]
\includegraphics[width=0.9\columnwidth]{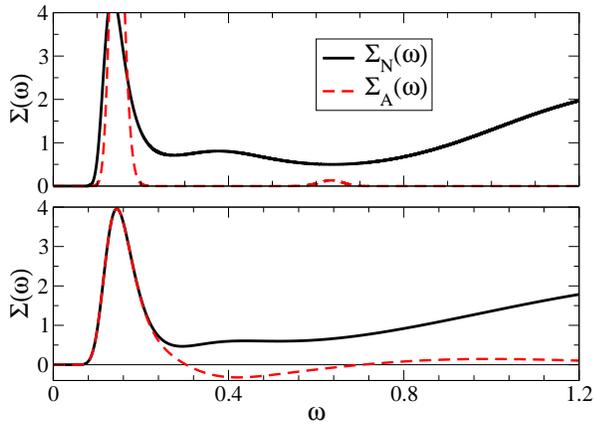}
\caption{Imaginary part of the normal ($\tau_0$,  black solid line) and anomalous ($\tau_1$, red dashed line)  components of the self energy calculated for particle-hole symmetric Hubbard model at $n=1$, $U=5.8,$ and $T=t/60$. Upper panel: self energies obtained by directly continuing Matsubara axis self energy. Lower panel: self energies  obtained by continuing $\Sigma_{\pm} = \Sigma_N \pm \Sigma_A$.}
\label{sigs}
\end{figure}

\begin{figure}[htb]
\includegraphics[width=0.9\columnwidth]{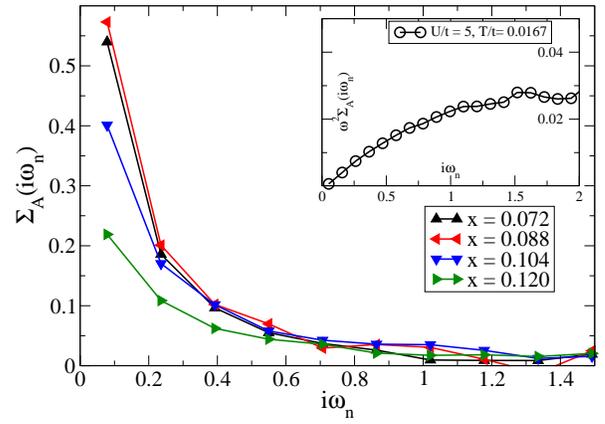}
\caption{Main panel: Matsubara frequency dependence of anomalous term in electron self energy computed for the $8$-site DCA approximation at $T=t/40,$ $U=7t,$ and densities indicated. %Left inset: comparison of frequency dependence of anomalous self energy for $4$ and $8$-site cluster approximations computed at $U=6t, T=t/40$ and $x = 0.08.$ Right 
Inset: product $\omega^2 \Sigma_A(i\omega_n)$ demonstrates $1/\omega_n^2$ decay of anomalous self energy for $U=5t$ and $x=0$.}
\label{FigFreq}
\end{figure}

These issues may be addressed in part by consideration of the particle-hole symmetric case (half filling, $t^{'}=0$) where the impurity-model Nambu Green function is diagonalized at all frequencies by the combinations $G_\pm=G_N\pm G_A$. In this case we can continue $\Sigma_\pm=\Sigma_N\pm\Sigma_A$. The corresponding  spectral functions are positive definite so the continuation is mathematically non-problematic.  We may then  reconstruct the normal and anomalous components from the sum and difference $\Sigma_+\pm\Sigma_-$. Results obtained for $n=1$ and $U=5.8t$ are shown in the lower panels of Fig.~\ref{sigs}.  At frequencies $\omega \gtrsim 0.2$, the inevitable errors in the continuations (in particular the small amplitude long period oscillations mentioned above) lead to a small oscillating component which gives rise to regions in which  $\Sigma_A^{''}(\omega>0)<0$. Because these are non-systematic with continuation method, variation of $U$ and temperature so we ascribe them to continuation errors.  We conclude that within our numerical and continuation uncertainties, $\Sigma_A(\omega>0)$ is positive definite.  More importantly, the basic structure (a large pole at $\omega\sim 0.15t$, of comparable strength in the normal and superconducting channels) is found in both methods. The dominance of this pole can be seen directly  from the Matsubara axis data, without continuation. Fig.~\ref{FigFreq} shows the Matsubara axis frequency dependence of the anomalous term in the electron self energy. We see that the behavior is very close to the $\Sigma_A\sim 1/(\omega_n^2+\omega_0^2)$.

%In these continuations the pole strengths are comparable (integration, not shown, over the frequency range $0.1<\omega<0.2$ gives $0.32t^2$ for both curves)  but we see that the pole positions are not quite identical $\Omega_P^N\neq \Omega_P^A$ and   the widths are substantially  different with the pole in the anomalous self energy being narrower.  However, not all of the differences between the continued self energies are physical. In particular, the positivity of the density of states (imaginary part of trace of Nambu Green function) implies a delicate relation between the normal and anomalous  self energies which may be violated by the inevitable errors of analytical continuation. In fact the   density of states reconstructed from the self energies shown in Fig.~\ref{sigs}  has an unphysical sign change in the vicinity of the pole frequency, arising mathematically from the slight difference in pole position combined with the relative narrowness of the peak in $\Sigma^{''}_A$.  This problem does not occur in standard phonon-mediated superconductors because the pole strength in  anomalous part of the self energy is small and the broadening is large. 

%We have not found a reliable  procedure for ensuring that  analytical continuation of the self energies  respects the relation between $\Sigma_N$ and $\Sigma_A$ implied by the positivity of the density of states, but in 

%The same phenomena can be seen directly from the Matsubara axis data, without continuation. Fig.~\ref{FigFreq} shows the Matsubara axis frequency dependence of the anomalous term in the electron self energy. We see that the behavior is very close to the $\Sigma_A\sim 1/(\omega_n^2+\omega_0^2)$

\begin{figure}[t]
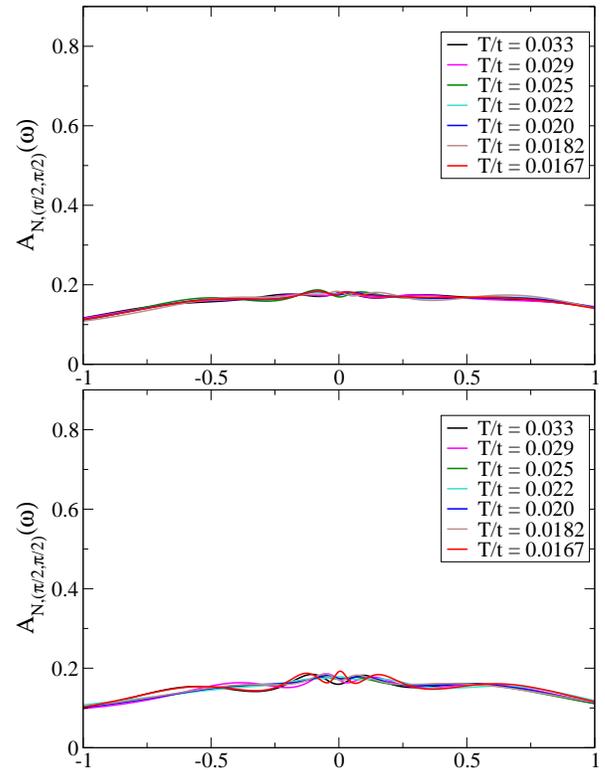

\includegraphics[width=0.9\columnwidth]{figures/NodalSpectralFunctionMum08}
\includegraphics[width=0.9\columnwidth]{figures/NodalSpectralFunctionMum04}
\caption{Spectral function $A_{\pi/2,\pi/2}(\omega)$ obtained by analytically continuing the normal part of the Green's function of the nodal momentum patch. Compare to Fig.~2 of the main text.}
\label{fig:nodalspectra}
\end{figure}

For completeness we present in Fig.~\ref{fig:nodalspectra} the spectral functions integrated over the nodal patch, obtained analogously as Fig. 2 in the main text, for two doping $x=0.076$ (OD, upper panel) and $x=0.034$ (UD, lower panel). This data shows a completely featureless spectral function and illustrates the metallic behavior at the node, consistent with data in the normal state (see e.g. Refs.~\onlinecite{Gull10_clustercompare,Gull09_8site,Lin10}).

\bibliographystyle{apsrev}
\bibliography{refs_shortened}